\documentclass{ws-ijmpe}
\usepackage{graphicx}

\begin{document}
\markboth{B. R. Barrett, D. M. Cardamone, and C. A. Stafford}{Superdeformed Nuclei}

\title{Exactly Solvable Model for the Decay of Superdeformed Nuclei}

\author{\footnotesize B. R. Barrett, D. M. Cardamone, and C. A. Stafford}

\address{Department of Physics, P. O. Box 210081, University of Arizona, 1118 E. 4th St.\\
Tucson, Arizona 85719, United States of America}

\maketitle

%\begin{history}
%\received{(received date)}
%\end{history}

\begin{abstract}
The history and importance of superdeformation in nuclei is briefly
discussed. A simple two-level model is then employed to obtain an elegant
expression for the branching ratio for the decay via the E1 process in the
normal-deformed band of superdeformed nuclei. From this expression, the
spreading width $\Gamma^\downarrow$ for superdeformed decay is found to be
determined completely by experimentally known quantities. The accuracy of the
two-level approximation is verified by considering the effects of other
normal-deformed states. Furthermore, by using a statistical model of the
energy levels in the normal-deformed well, we can obtain a probabilistic expression for the tunneling matrix element $V$.
\end{abstract}

\section{Superdeformed Nuclear Decay}

Superdeformation is one of the most interesting examples
of collective phenomena in atomic nuclei. Since its original experimental
observation in 1986,\cite{twin86} the properties of these high-spin
rotational bands have fascinated experimentalists and theoreticians alike. When
combined with precisely measured branching ratios and decay rates, a thorough
theoretical understanding of the mechanism by which superdeformed (SD) nuclei
are formed and then decay
into normal-deformed (ND) bands promises to provide a window into nuclear structure
unlike any other.

One of the major barriers to using SD decay to understand nuclear structure
has been that there is no clear way to link the quantities measured in
experiment directly with those which might shed light on the internal dynamics of
the nucleus. Despite enormous strides made by experimentalists to measure
observables precisely,\cite{lauritsen,krucken,wilson03} theorists have failed to reach a consensus on
just what to do with these data. Ideally, we require a model which, while accounting for
the rich physics of the SD nucleus, is also as simple as possible to allow
for easy extraction of quantities of interest to theory.

\begin{figure}
\includegraphics[keepaspectratio=true,width=\columnwidth]{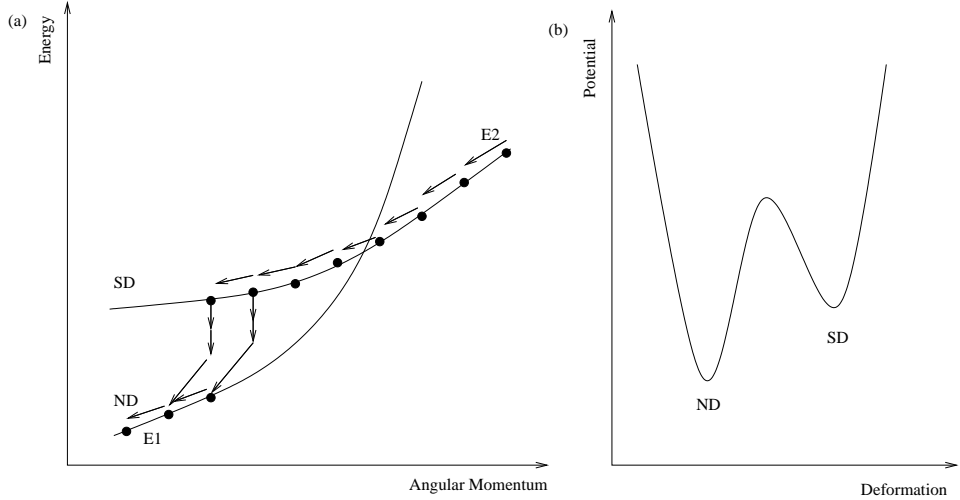}
\caption{(a) Schematic diagram of the decay-out process. The SD band loses
  essentially all its
  strength to an ND band over the course of a few
  states. (b) The problem is commonly modeled with a double-well potential
  in deformation space. The shape of the potential is a function of the
  angular momentum of the nucleus.}
\label{decay}
\end{figure}

The typical SD nuclear experiment\cite{twin86,lauritsen,krucken,wilson03} creates nuclei in a high
angular momentum state in the SD potential well. These nuclei then
lose rotational energy by E2 transitions, eventually reaching a low enough
angular momentum that it becomes energetically favorable to decay to states of
the same angular momentum in the ND well. As each SD nucleus continues to
decay, more strength moves into the
energetically preferred ND band rather than continuing down the band of SD
states. In practice, most of the decay-out of the SD band happens over the
space of only two or three SD states, after which essentially all the strength has moved into the ND band. For a schematic diagram of this process, see Figure
\ref{decay}(a).

It is customary to model the decay process as a double well problem, as in
Figure \ref{decay}(b). The
nucleus is a quantum mechanical system moving in a spin-dependant potential in
deformation space. The potential consists of two wells, and the eigenstates of
each well in the absence of the other make up the pure SD and ND bands. In
such models, it is the shape of the potential at various angular momenta which contains information about
the underlying nuclear structure. Thus, an important goal for theories of this
type is that they provide a method to relate the potential to experimental observables.

\section{Two-State Model}
The two-state model for SD nuclear decay \cite{sb99} is an exactly solvable
approximation to the decay-out scheme outlined in the previous section. The
key assumption is that only one ND level participates significantly in the
decay of a particular
SD level of the same angular momentum. This reduces the problem to the familiar two-level mixing scenario
of introductory quantum  mechanics. The great advantage of this approach is that the two-state model
rigorously yields simple formulae connecting experimental observations to the
physics of the potential barrier.

\begin{figure}
\begin{center}
\includegraphics[keepaspectratio=true,width=7cm]{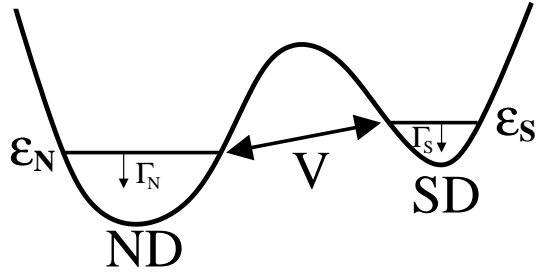}
\end{center}
\caption{Schematic diagram of the two-level model for SD nuclear
  decay. $\varepsilon_S$ and $\varepsilon_N$ are the energies of the unmixed SD
  and ND states, respectively; $\Gamma_S$ and $\Gamma_N$ are energy widths due
  to interaction with the electromagnetic field; and $V$ is the matrix element
  for tunneling through the barrier.}
\label{2well}
\end{figure}

Keeping only the ND level of the same angular momentum nearest in energy to
the decaying SD state, one can simplify
the system to that shown in Figure \ref{2well}. The dynamics of the
model are thus completely determined by four parameters: the difference in the
unmixed levels' energies
$\Delta=\varepsilon_N-\varepsilon_S$, the energy widths for coupling to the
electromagnetic environment $\Gamma_S$ and $\Gamma_N$, and the tunneling
matrix element $V$. A fifth parameter, the mean level spacing in the ND
band $D_N$, can be taken as the unit of energy.

In the absence of the tunneling parameter $V$, the system reduces to two
isolated Breit-Wigner resonances, and the retarded Green's function in the
energy domain is simply
\begin{equation}
G_0(E)=\left( \begin{array}{cc}
\frac{1}{E+i\Gamma_S/2}&0\\
0&\frac{1}{E-\Delta+i\Gamma_N/2}\end{array}\right)
\end{equation}
in the $S,N$ basis. Coupling between the two levels is given by the simple
perturbation matrix
\begin{equation}
\hat{V}=\left(\begin{array}{cc}
0&V\\
V&0\\
\end{array}\right).
\end{equation}
The full Green's function is given exactly to all orders in
$\hat{V}$ by Dyson's Equation. Its inverse is
\begin{equation}
G^{-1}(E)=G_0^{-1}(E)-\hat{V}=\left(\begin{array}{cc}
E+i\Gamma_S/2&-V\\
-V&E-\Delta+i\Gamma_N/2\end{array}\right).
\end{equation}

The retarded Green's function contains all information about the time evolution
of the system. We are specifically interested in the branching ratios, which
are given by Parseval's Theorem to be
\begin{equation}
\label{parseval}
F_i=\Gamma_i\int^\infty_{-\infty}\frac{dE}{2\pi}|G_{iS}(E)|^2,
\end{equation}
where $i$ is either $S$ or $N$. The ND branching ratio follows:
\begin{equation}
\label{fn}
F_N=\frac{\Gamma_N\Gamma^\downarrow/(\Gamma_N+\Gamma^\downarrow)}{\Gamma_S+\Gamma_N\Gamma^\downarrow/(\Gamma_N+\Gamma^\downarrow)},
\end{equation}
a result we first noted in Ref.\ \refcite{csb03}. Here $\Gamma^\downarrow$ is the
spreading width for tunneling through the barrier:
\begin{equation}
\label{gdown}
\Gamma^\downarrow=\frac{2\overline{\Gamma}V^2}{\Delta^2+\overline{\Gamma}^2},
\end{equation}
where $\overline{\Gamma}\equiv\frac{\Gamma_N+\Gamma_S}{2}$. 

In this case, Equation (\ref{gdown}) is the exact result of Eq.\
(\ref{parseval}), and it is also what one expects from correct
application of Fermi's Golden Rule.\cite{sb99} We emphasize that, in general,
$\Gamma^\downarrow\neq\langle\Gamma^\downarrow\rangle$, the ensemble average
of $\Gamma^\downarrow$. In much previous literature (e.g., Ref.\ \refcite{vigezzi}), it was
assumed that the two are equivalent, giving erroneous results. In fact, they are
equal only in the continuum limit of overlapping resonances.\cite{sb99} This
is
clearly not an appropriate approximation to most nuclei of interest, in which
$D_N$ exceeds $\Gamma_N$ by several orders of magnitude. Indeed, in real
nuclei the
discrepancy between $\Gamma^\downarrow$ and its ensemble average can be as
much as three to four orders of magnitude %\cite{csb03}.
(see Table 1).

\begin{table}[pt]
\tbl{Values of $\Gamma^\downarrow$ and $\langle V\rangle$ for various
  SD decays in the $A\approx 150$ and $A\approx 190$ mass
  regions. $\langle\Gamma^\downarrow\rangle$ from Ref.\
  \protect\refcite{lauritsen} is tabulated for comparison. Values for $F_N$,
  $\Gamma_S$, $\Gamma_N$, and $D_N$ are also from Ref.\
  \protect\refcite{lauritsen}. $\Gamma^\downarrow$ is calculated from Eq.\
  (\ref{gdownresult}), either directly or, in the case of $^{152}\mathrm{Dy}(26)$, statistically, as discussed
  in the text, while $\langle V\rangle$ is given by Eq.\
  (\ref{vave}). Angular momenta for the decaying SD states are given
  in parentheses.}
{\begin{tabular}{ccccc|cc|c}\toprule
& $F_N$ & $\Gamma_S$ & $\Gamma_N$ & $D_N$ & $\Gamma^\downarrow$ & $\langle V\rangle$ &
  $\langle\Gamma^\downarrow\rangle$\\
& $=P_{out}$ & (meV) & (meV) & (eV) & (meV) & (eV) & (meV)\\ \colrule
$^{152}\mathrm{Dy}(28)$ & 0.40 & 10.0 & 17 & 220 & 11 & 35 & 41 000 \\
$^{152}\mathrm{Dy}(26)$ & 0.81 & 7.0 & 17 & 194 & 78 & 87 & 220 000 \\
$^{194}\mathrm{Hg}(12)$ & 0.40 & 0.108 & 21 & 344 & 0.072 & 5.0 & 560 \\
$^{194}\mathrm{Hg}(10)$ & 0.97 & 0.046 & 20 & 493 & 1.6 & 35 & 37 000 \\ \botrule
\end{tabular}}
\end{table}

Equation (\ref{fn}) demonstrates the power of using a simple, intuitively
understandable model. We can instantly see that the result is correct, since
it can be read as the branching ratio of a two-stage decay problem. That is,
$F_N$ is the probability for the nucleus to tunnel through the barrier \emph{and then} to decay down the ND band. Furthermore, it is eminently useful:
we see that, given a set of experimental results, $\Gamma^\downarrow$ is
known. Inverting Eq.\ (\ref{fn}) we find
\begin{equation}
\label{gdownresult}
\Gamma^\downarrow=\frac{F_N\Gamma_N\Gamma_S}{\Gamma_N-F_N(\Gamma_N+\Gamma_S)},
\end{equation}
so that the spreading width is fixed by experimental results. The recent work
of Wilson and Davidson\cite{wilson04} has shown how to relate
$\Gamma^\downarrow$ (Eq.\ (\ref{gdownresult})) to the barrier height by assuming it is equivalent to a fusion-like tunneling rate.

Since $\Gamma^\downarrow$ is a positive definite energy width, Eq.\
(\ref{gdownresult}) implies a lower bound in the two-level model for
the decay width due to E1 transitions in the ND band,
\begin{equation}
\label{gnineq}
\Gamma_N>\frac{F_N\Gamma_S}{1-F_N}.
\end{equation}
This prediction of the two-level model is especially significant due
to the relatively high uncertainty in current values for $\Gamma_N$.

The estimated value\cite{lauritsen} of $\Gamma_N$ for
$^{152}\mathrm{Dy}(26)$ violates inequality (\ref{gnineq}). This could
indicate a breakdown of the two-level approximation. However, such a
conclusion would be precipitous due to the large uncertainty in the
current best estimate of $\Gamma_N$, which could be off by a factor of
two or more. An estimate of $\Gamma^\downarrow$ for this case can be
obtained in the two-level model by cutting off the normal distribution
for $\Gamma_N$ with inequality (\ref{gnineq}). The median value of
$\Gamma^\downarrow$ so estimated is presented in Table 1, assuming the
standard deviation $\sigma_{\Gamma_N}=\Gamma_N=17\mathrm{meV}$.

\section{Gaussian Orthogonal Ensemble}
We have seen that the two-level model gives an expression for the spreading
width completely in terms of experimentally measurable
quantities. Since $\Gamma^\downarrow$ is clearly related to the potential
barrier between the two wells, this is a great stride toward the goal
laid out in the first section, namely to find a simple relationship between
observables and one or more parameters related directly to the shape of the
potential.

The spreading or decay width $\Gamma^\downarrow$, however, is a rate, and less directly related to the shape
of the potential than $V$, the tunneling matrix element itself. The
problem in relating $V$ directly to experiment is clear from Eq.\ (\ref{gdown}):
we would need to know the value of $\Delta$, which is unfortunately
not accessible with current experimental techniques.
Lacking a complete solution of the nuclear structure problem, we
derive a probabilistic theory for $V$, with its
roots in the Gaussian Orthogonal Ensemble (GOE).

The GOE is a ``structureless'' assumption for the level distribution of the
unmixed ND band, in which we assume only that the levels are eigenstates
of some Hermitian Hamiltonian with time reversal and SO(3) symmetry. The
states are characterized by the Wigner surmise,\cite{mehta91} a probability distribution for
the spacing $sD_N$ between levels:
\begin{equation}
P(s)=\frac{\pi}{2}se^{-\pi s^2/4}.
\end{equation}

Since we are interested in finding a probability distribution for unmixed
levels, the SD and ND spectra are uncorrelated. Thus, given $s$, we have a
rectangular probability density function for $\Delta$
\begin{equation}
\mathcal{P}_s(\Delta)=\frac{1}{sD_N}\Theta\left(\frac{s}{2}-\frac{|\Delta|}{D_N}\right).
\end{equation}
The Heaviside step function $\Theta$ specifies that $\Delta$ is the
detuning of the \emph{nearest} neighbor.

Application of basic probability theory leads us to the density function for
$\Delta$:
\begin{equation}
\mathcal{P}(\Delta)=\int_0^\infty\mathcal{P}_s(\Delta)P(s)ds=\frac{\pi}{2D_N}\mathrm{erfc}\left(\sqrt{\pi}\frac{|\Delta|}{D_N}\right).
\end{equation}
Here $\mathrm{erfc}(x)$ is the complementary error function of $x$. The
mean of this distribution is $D_N/4$.

From Eq.\ (\ref{gdown}) we note that
\begin{equation}
\label{delta}
|\Delta|=\sqrt{\frac{2\overline{\Gamma}}{\Gamma^\downarrow}\left(V^2-\frac{\Gamma^\downarrow\overline{\Gamma}}{2}\right)},
\end{equation}
which implies
\begin{equation}
V\ge V_{min}=\sqrt{\frac{\Gamma^\downarrow\overline{\Gamma}}{2}}.
\end{equation}
This allows calculation of the desired probability density function for $V$:
\begin{equation}
\label{vprob}
\mathcal{P}(V\ge V_{min})=2\mathcal{P}(\Delta)\left|\frac{d\Delta}{dV}\right|=\frac{2\overline{\Gamma}V\pi}{\Gamma^\downarrow|\Delta|D_N}\mathrm{erfc}\left(\sqrt{\pi}\frac{|\Delta|}{D_N}\right),
\end{equation}
where $\Delta$ is given as a function of $V$ by Eq.\ (\ref{delta}).

We have arrived at a probability density function for $V$ solely in terms of
experimentally measurable parameters. The mean of this distribution is
\begin{equation}
\label{vave}
\langle
V\rangle=\sqrt{\frac{\Gamma^\downarrow}{2\overline{\Gamma}}}\left[\frac{D_N}{4}+\mathcal{O}\left(\frac{\overline{\Gamma}^2}{D_N}\right)\right].
\end{equation}
Values of $\langle V\rangle$ for specific SD decays are given in Table 1. In general, we can expect that $\langle V\rangle$ will be a typical value of
$V$ for nuclei measured in the laboratory, since the standard deviation of the
distribution (\ref{vprob}) is comparable to the mean (\ref{vave}).

Equation (\ref{vave}) is a profound success of the two-level model. In fact,
even by including more levels, a probabilistic statement like this represents
the \emph{most} information one can have about $V$ with the current types of experiments.

\section{Beyond the Two-level Model}
In the previous two sections, we solved an approximate model exactly. In reality,
of course, the ND band is semi-infinite, starting at the bandhead and
continuing upward. We now address the validity of the two-level approximation.

The simplest way to address this issue is to add a second ND
level.\cite{csb03} The branching ratio is now calculated in a three-dimensional
Hilbert space, in exactly the same way as before. The inverse Green's function
is
\begin{equation}
G^{-1}(E)=\left(\begin{array}{ccc}
E+i\Gamma_S/2&-V_1&-V_2\\
-V_1&E-\Delta_1+i\Gamma_N/2&0\\
-V_2&0&E-\Delta_2+i\Gamma_N/2\end{array}\right),
\end{equation}
where the subscripts $1$ and $2$ pertain, respectively, to the nearest and next-nearest ND
states to the decaying SD with the same angular momentum.

\begin{figure}
\includegraphics[keepaspectratio=true,width=\columnwidth]{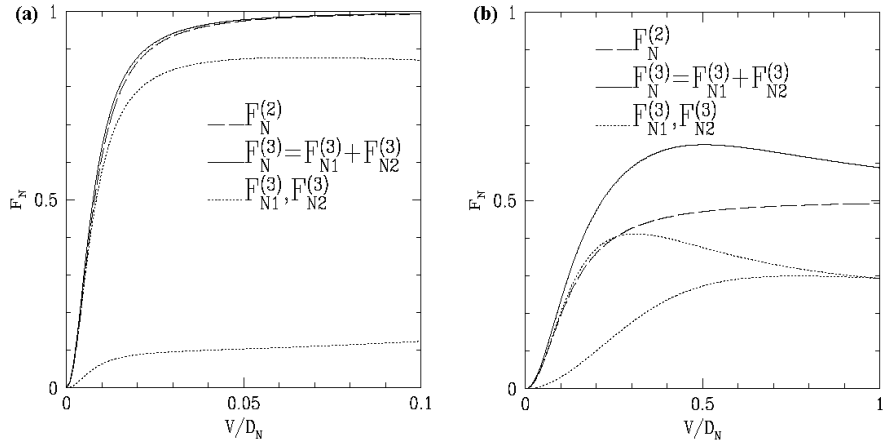}
\caption{Numerical results comparing the two- and three-state total branching
  ratios. Superscripts indicate the model used whereas subscripts represent
  the nearest (1) and next-nearest (2) neighbor in the three-level case. (a)
  $A\approx 190$ mass region, as determined by
  $\Gamma_S/\Gamma_N=10^{-3}$. (b) $A\approx 150$ mass region with
  $\Gamma_S=\Gamma_N$. In
  both cases $\overline{\Gamma}/D_N=10^{-4}$. The levels were taken at their
  mean detunings within the GOE under the
  assumption that they lie on either side of the SD state, and in both cases $V_1=V_2=V$.}
\label{3state}
\end{figure}

It is straightforward to obtain an analytic result for the three-level system
by applying Eq.\ (\ref{parseval}), but since we desire to know how changed the
total branching ratios are from the two-state results over a range of
parameters, it is perhaps more useful to calculate them numerically. Figure
\ref{3state} shows the results for the $A\approx 190$ and $A\approx 150$ mass
regions.

We note that the figure shows remarkably good agreement between the
two- and three-level total branching ratios in the mass-190 region,
particularly for values of $V$ that are likely according to Eq.\
(\ref{vave}). In the mass-150 region, as for $^{152}Dy$, they may disagree
by as much as 25\% ($I=28$) to 40\% ($I=26$) due to the larger value
of the coupling matrix element $V$ and the smaller level spacing,
\emph{i.\ e.}, higher density of states. This supports the validity of
the two-level approximation for the mass-190 region, in the sense that
the addition of the third level has not affected the physically
measured quantities too significantly. In the mass-150 region, the
three-level approximation yields a significant correction to the
two-state branching ratios for values of $V/D_N\gtrsim 0.15$.

Further support for the two-level approximation is given by Ref.\
\refcite{dzyublik03}, in which Dzyublik and Utyuzh exactly solve the problem in
the approximation of an infinite (in both directions), equally spaced ND
band. Their results also show remarkably strong agreement with the two-level
approximation for typical values of $\Delta$.

\section{Conclusions}
Our three-state results, together with the results of Ref.\ \refcite{dzyublik03},
demonstrate that the two-state model is sufficient to describe the dominant
decay-out process of SD nuclei. Within the two-state model, we have shown
that the decay out of an SD level via the E1 process in the ND band is a two-step
process, whose branching ratio (\ref{fn}) is expressed in terms of three
measurable rates, $\Gamma_S$, $\Gamma_N$, and $\Gamma^\downarrow$. We have also
shown how to determine the tunneling matrix element $V$ (Eqs.\ (\ref{vprob})
and (\ref{vave})) from the measured values of $\Gamma^\downarrow$ and a
statistical model of the ND band. It is hoped that these results will help to
clarify the nature of the decay-out process in SD nuclei, and that their
elegance will inspire further promising studies such as Ref.\ \refcite{wilson04}.

\section*{Acknowledgments}
We acknowledge support from NSF Grant No.\
PHY-0210750. B. R. B. acknowledges partial support from NSF Grant
Nos.\ PHY-0070858 and PHY-0244389. We also thank the Institute of
Nuclear Theory at the University of Washington for its hospitality and
the Department of Energy for partial support during the formulation of
this work.

B. R. B. thanks the organizers and sponsors for making \emph{Blueprints for the Nucleus} possible,
with special thanks to the Feza Gursey Institute for hosting the conference.

\end{document}